\documentclass{article}
\usepackage{spconf,amsmath,graphicx,hyperref}
\usepackage{siunitx}
\usepackage{comment}
\usepackage{bm} % for bold math symbols (incl. Greek  letters) with \bm{}
\usepackage{booktabs}
\usepackage{array}     % for p{} column formatting
\usepackage{hyperref}

\title{Room Impulse Response Prediction with Neural Networks: From Energy Decay Curves to Perceptual Validation}
\name{Imran Muhammad, Gerald Schuller}
\address{Technische Universität Ilmenau, Germany}
\begin{document}
%\ninept
%
\maketitle

\begin{abstract}
Prediction of room impulse responses (RIRs) is essential for room acoustics, spatial audio, and immersive applications, yet conventional simulations and measurements remain computationally expensive and time-consuming. This work proposes a neural network framework that predicts energy decay curves (EDCs) from room dimensions, material absorption coefficients, and source–receiver positions, and reconstructs corresponding RIRs via reverse-differentiation. A large training dataset was generated using room acoustic simulations with realistic geometries, frequency-dependent absorption, and diverse source–receiver configurations. Objective evaluation employed root mean squared error (RMSE) and a custom loss for EDCs, as well as correlation, mean squared error (MSE), spectral similarity for reconstructed RIRs. Perceptual validation through a MUSHRA listening test confirmed no significant perceptual differences between predicted and reference RIRs. The results demonstrate that the proposed framework provides accurate and perceptually reliable RIR predictions, offering a scalable solution for practical acoustic modeling and audio rendering applications.
\end{abstract}
\begin{keywords}
Room Acoustics, Neural Network, Energy Decay Curves, Auralization, Virtual Reality
\end{keywords}
\section{Introduction}
\label{sec:intro}

The room impulse response plays a vital role in room acoustics, spatial audio rendering, and immersive virtual environments. Standard room acoustic parameters, which provide quantitative insight into acoustic behavior of the closed space, are often derived from RIRs or EDCs, which describe how sound energy evolves over time after an acoustic excitation \cite{kuttruff2009room}.
Classical room-acoustic modeling relies on computationally expensive simulations (e.g., ray tracing (RT), image-source method (ISM), and wave-based simulations) \cite{vorlander2020auralization} or time-consuming on-site measurements, which limit scalability and flexibility. Fast acoustics software such as pysound \cite {schissler2011gsound}, ITA Geometrical Acoustics \cite{ITAGeometricalAcoustics} openRay \cite{openRay}, EVERTims \cite{Laine2009}, GSoundSIR \cite{GSoundSIR} and Pyroomacoustics \cite {8461310} enable an efficient alternative to physical measurements. However, these methods are based on approximations and struggle to capture wave-based phenomena accurately, particularly at low frequencies. Wave-based solvers offer greater accuracy but are often computationally intensive and unsuitable for real-time applications. This trade-off between speed and accuracy presents an opportunity for deep learning or data-driven approaches. Crucially, such models should be trained on high-quality data, ideally derived from physical measurements or wave-based simulations, to ensure reliability.

Data-driven approaches have grown rapidly in the last few years, targeting either (i) prediction of room parameters from descriptors of the space, \cite{foy2021mean} (ii) direct estimation or completion of room impulse responses (RIRs), \cite{lin2025deep, karakonstantis2024physics} or (iii) downstream dereverberation and robust ASR. Early neural predictors focused on mapping geometric and material features to scalar reverberation metrics (e.g., T60, clarity), demonstrating that learned regressors can approximate or outperform simplified formula-based estimates in complex rooms \cite{meng2023predicting, 10.1121/10.0013575, 10.1121/10.0013416,9286412}. Beyond scalar metrics, several works aim to reconstruct full RIRs. Deep RIR completion formulates RIR completion as a lightweight learning task: given partial or bandlimited information, a neural model infers broadband RIRs, evaluating with EDF/EDC errors, T60 MSE, and DRR MSE across multiple datasets \cite{kim2023generative, masztalski2020storirstochasticroomimpulse, lin2025deep}. These advancements highlight the effectiveness of deep learning in capturing the intricate temporal and spatial characteristics of room acoustics, providing efficient and scalable alternatives to conventional physics-based modeling techniques.

We propose a neural network–based framework for estimating RIRs by first predicting EDCs directly from input parameters describing the acoustic environment. The input features include room dimensions, material absorption properties, and source–receiver positions, which collectively define the acoustical behavior of a given space. In contrast, RIRs are high-dimensional signals with complex temporal and spectral details, making direct prediction more challenging and error-prone. Since EDCs also serve as the basis for computing important acoustic parameters \cite{vorlander2020auralization}, this strategy simplifies the modeling task while preserving the essential information required for an accurate room acoustics analysis. This framework centers on a neural model that maps room geometry and material properties to EDCs. Using EDCs as an intermediate representation captures temporal energy decay, which are then used to reconstruct full RIRs through reverse differentiation and polarity assignment. This ensures consistency of temporal and spectral features while leveraging the stability of EDC modeling. Training relied on a large simulated dataset incorporating varied geometries, frequency-dependent absorption, and source–receiver configurations to enable generalization. Evaluation combined objective and perceptual measures: root mean square error (RMSE) and a custom temporal-decay loss for EDCs, whereas correlation, RIR-MSE, and spectral MSE for RIRs provided a comprehensive assessment of accuracy, while perceptually motivated approaches such as psychoacoustic losses have also been explored in related work \cite{10943044}.

Finally, perceptual validation was conducted through a MUSHRA listening test. The results demonstrated that listeners perceived no significant differences between the predicted and reference RIR conditions, highlighting the effectiveness of the proposed method in delivering perceptually convincing outcomes. This confirms that the framework not only achieves strong numerical performance but also satisfies perceptual standards, making it a promising tool for applications in virtual acoustics, audio rendering, and acoustic design.

\section{Methodology}
\label{sec:methodology}
We propose a deep learning framework based on Long Short-Term Memory (LSTM) neural networks to predict energy decay curves (EDCs) directly from room geometries and wall absorption characteristics. LSTMs are well suited for sequential modeling as they capture long-term temporal dependencies and mitigate vanishing gradients, a limitation of standard recurrent neural networks (RNNs) \cite{hochreiter1997long, bengio1994learning}. Compared to convolutional neural networks (CNNs), which focus on spatial pattern recognition, LSTMs are more effective for modeling the temporal decay of EDCs. While Transformer architectures achieve strong sequence modeling performance, their computational demands and data requirements make them less practical in this setting \cite{vaswani2017attention}. LSTMs therefore offer a robust and efficient solution for predicting EDCs from room features. The model is trained on synthetic dataset covering varied room dimensions, source–receiver placements, and surface absorption properties. Input features are normalized and structured for LSTM model, and training is guided by a custom loss function (Sec.~\ref{subsec:LSTM Model Architecture}) tailored for accurate EDC prediction.

\subsection{Dataset and Room Parameters}
\label{ssec:dataset}
A dataset of $6000$ simulated shoebox rooms was generated using the Pyroomacoustics library \cite{scheibler2018pyroomacoustics}, with each room defined by length (\SI{}{L}), width (\SI{}{L}), height (\SI{}{H}), source/receiver 3D positions (\SI{}{X}, \SI{}{Y}, \SI{}{Z}), and frequency-dependent absorption coefficients averaged across surfaces \cite{8461310}. Room dimensions, positions, and material properties were systematically varied (Table \ref{tab:simulated_room_ranges}) to span realistic acoustic conditions. Pyroomacoustics combines the image source method (ISM) and ray tracing to approximate high-frequency sound propagation, efficiently generating RIRs and EDCs at \SI{48}{\kilo\Hz}. While limited by its geometric assumptions (ignoring wave effects), this approach covers a broad range of reverberation times (T60), from nearly anechoic (\SI{0.1}{s}) to highly reverberant (\SI{3.0}{s}).

\begin{table}[t]
\centering
\caption{Range of simulated room features used as input parameters for LSTM model training}
\vspace{2pt} 
\label{tab:simulated_room_ranges}
\begin{tabular}{@{}>{\raggedright\arraybackslash}p{0.5\linewidth}c@{}}
\toprule
\textbf{Parameter} & \textbf{Range / Values} \\
\midrule
Room Dimensions (L×W×H)& 3-6 m × 3-6 m × 2.5-4 m\\
Source-Receiver Distances& 1 - 4 m\\
Wall Absorptions Range& 0.14 - 0.65 (125 - 8000 Hz)\\
No. of Room Configurations& 6000\\
No. Material Types& 10\\
\end{tabular}
\end{table}

\begin{comment}
\begin{figure}
    \centering
    \includegraphics[width=1\linewidth]{histogramRT20_1.jpg}
    \caption{Histogram showing the distribution of $T60$ values measured across the room dataset}
    \label{fig:RTHist}
    \vspace{-0.5cm}
\end{figure}
\end{comment}

Input features were scaled using \textit{MinMax} normalization to $[0,1]$ for consistent training dynamics. Although EDCs already fall within this range, normalization was applied as a standard deep learning practice. Data were reshaped into $[N,1,16]$, where $N$ is the number of rooms and $16$ the features per room, to match the input format required by the LSTM architecture (batch\_size, sequence\_length, feature\_dimension). Each room sample is thus represented as a one-step sequence of $16$ values, enabling the LSTM to capture potential inter-dependencies. For evaluation, the dataset was split into $60\%$ training, $20\%$ validation, and $20\%$ testing.

\subsection{LSTM Model Architecture}
\label{subsec:LSTM Model Architecture}
The LSTM architecture was implemented in PyTorch with $128$ hidden units to capture dependencies among input features via memory cells and gating mechanisms. To reduce overfitting, a $30\%$ dropout was applied after the LSTM. This was followed by a dense layer of $2048$ \textit{ReLU}-activated neurons for non-linear feature learning, with another dropout for regularization. The output layer matched the EDC sequence length, using a linear activation for continuous prediction. Training used the Adam optimizer ($LR=0.001$) and mean squared error (MSE) with an additional custom loss. Early stopping (patience of $10$) prevented overfitting and improved efficiency. Although training was capped at $200$ epochs, convergence often required fewer iterations.

We designed a composite loss function that integrates two complementary terms: EDCs fidelity, RIR fidelity. Let $\mathbf{y} \in \mathbf{R}^T$ denote the ground-truth (target) EDC of length $T$, and $\hat{\mathbf{y}} \in \mathbf{R}^T$ the predicted EDC. The corresponding RIR signals are obtained by using RSS approach in Section \ref{subsec:rirSynthesis}. The total loss is defined as a weighted sum of two terms.

\begin{equation}
\label{equation: eq1}
\mathcal{L} = \alpha \, \mathcal(\frac{1}{T}\sum_{t=1}^{T} \big( \hat{y}_t - y_t \big)^2) + \beta \, \mathcal(\frac{1}{T-1}\sum_{t=2}^{T} \big( \hat{r}_t - r_t \big)^2)
\end{equation}
Where $\alpha, \beta \geq 0$ are weighting coefficients and $\mathbf{r}_t$ denotes the RIR obtained from the ground-truth EDC, and $\hat{\mathbf{r}} t$ denotes the RIR obtained from the predicted EDC. We choose $\alpha = 1$ and $\beta = 0.5$. In the Eq. \ref{equation: eq1}, the EDC loss enforces accurate cumulative decay prediction, where as the RIR loss constrains the predicted EDC to yield a realistic RIR shape.

\subsection{Reconstruction of Room Impulse Responses from Energy Decay Curves}
\label{subsec:rirSynthesis}
For a given a room impulse response $h(n)$, its instantaneous energy is defined as
\begin{equation}
e(n) = h^2(n),
\end{equation}
and the cumulative decay curve (EDC) is computed in reverse time as
\begin{equation}
\text{EDC}[n] = \sum_{k=n}^{N-1} e[k],
\end{equation}
where $T$ is the signal length. Differentiating the EDC redistributes the energy over time:
\begin{equation}
d[n] = -\big(\text{EDC}[n+1] - \text{EDC}[n]\big),
\end{equation}
and the magnitude response of the RIR can be estimated as
\begin{equation}
h_{\text{mag}}[n] = \sqrt{\max(d[n], 0)}
\end{equation}

Since only the magnitude can be recovered from the EDC, the polarity/phase must be reconstructed using stochastic methods. Two strategies are considered: i) the \textit{random sign} method, where each sample is assigned an independent random polarity, and ii) the \textit{random sign-sticky} method, which enforces sign continuity between consecutive samples. In our implementation, continuity is controlled by a \textit{stickiness parameter}, set to $0.90$, meaning that each sample has a $90\%$ probability of retaining the polarity of the previous one and a $10\%$ chance of flipping. A higher stickiness value promotes smoother, more coherent low-frequency oscillations, which is beneficial for long-reverberant RIRs, while a lower value allows faster sign changes, enhancing the representation of short reverberation tails. Empirically, stickiness values in the range $0.7$--$0.95$ were found to yield stable and perceptually plausible reconstructions across a variety of reverberation times.

\section{Evaluation}
\label{sec:Evaluation}
After training, the model demonstrates strong predictive accuracy across unseen room configurations. The accuracy of the predicted EDCs are evaluated to verify the accuracy of the model. The predicted EDCs are then used to compute the room acoustical parameters such as reverberation time (T20), early decay time (EDT) and clarity (C50). These parameters are compared with those derived from the target EDCs to further access the reliability of the model.

\begin{comment}
\begin{figure}
    \centering
    %, height=6cm
    \includegraphics[width=1\linewidth]{EDCs_comparisons.png}
    \caption{Target and predicted EDCs for randomly selected room}
    \label{fig:edcplot}
    %\vspace{-0.5cm}
\end{figure}
\end{comment}

\subsection{Objective Evaluation of EDCs}
\label{sec:edcEvaluation}
\begin{figure}
    \centering
    \includegraphics[width=1\linewidth]{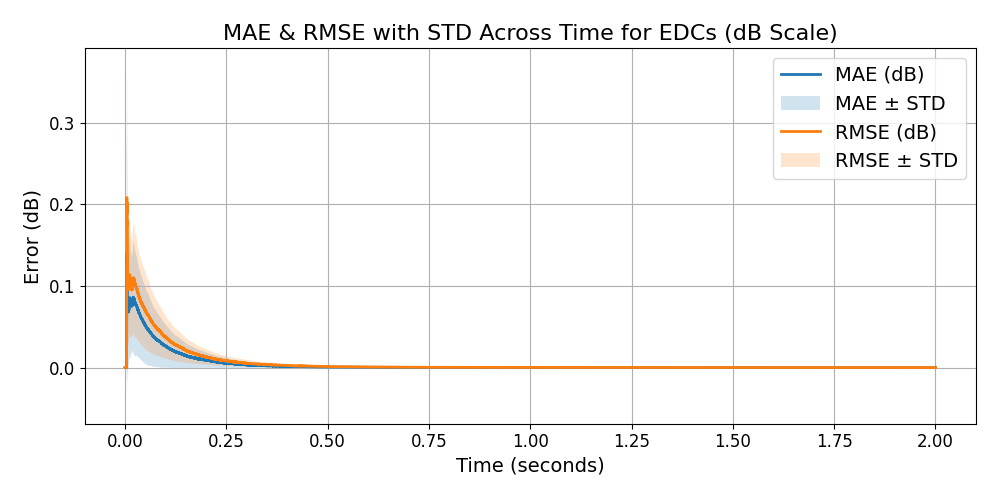}
    \caption{\textbf{MAE} and \textbf{RMSE} of the predicted EDCs, averaged over time across all room configurations}
    \label{fig:EDCError}
    \vspace{-0.5cm}
\end{figure}

To assess the model's predictive performance, qualitative comparisons were conducted between the target (i.e. simulated) and predicted EDCs. We reported the MAE and RMSE computed across all predicted EDCs, averaged over time. Additionally, we included the standard deviation to reflect variability across the dataset as shown in Figure~\ref{fig:EDCError} to provide a more comprehensive evaluation of model behavior. The graph reveal a strong agreement between the predicted and target EDC curves, suggesting that the model effectively captures the temporal decay characteristics of rooms. The predicted EDCs not only replicate the overall decay envelope but also capture fine-grained fluctuations in the decay profile.

\subsection{Objective Evaluation of Reconstructed RIRs}
\label{sec:rirEvaluation}
To objectively assess the quality of the reconstructed room impulse responses from the predicted energy decay curves, we computed several signal-level metrics, including the MSE between actual and predicted RIRs, correlation coefficients, and spectral MSE. RIRs were reconstructed from EDCs using the RSS approach. The objective results indicated that the reconstructed RIRs closely approximate the temporal and spectral characteristics of the ground-truth RIRs. From the four randomly selected predicted EDCs, the corresponding constructed RIRs showed stronger correlations ($0.37 - 0.52$) and lower spectral MSE (\SI{55}{} - \SI{58}{dB}). Table \ref{tab:acoustic_param_accuracy} summarizes the error metrics. These objective findings are corroborated by the perceptual evaluation conducted via a MUSHRA listening test, where the RSS-based reconstructed RIRs were rated significantly closer to the reference RIRs (sec \ref{subsec:mushra}). Overall, the analysis demonstrates that RIRs can be effectively reconstructed from predicted EDCs using the Random Sign-Sticky method, achieving both objective similarity and perceptual plausibility.

\begin{table}[ht]
\centering
\caption{Objective comparison of the four reconstructed RIRs using the RSS-approach. Metrics include RIR mean squared error (RIR-MSE), correlation coefficient, and spectral MSE.}
\vspace{4pt} 
\label{tab:rir_top4_single}
\begin{tabular}{c c c c}
\hline
No. & \textbf\textbf{RIR-MSE} & \textbf{Correlation} & \textbf{Spectral MSE (dB)} \\
\hline
1 & $1.20\times10^{-5}$ & 0.515 & 58.26 \\
2 & $1.22\times10^{-5}$ & 0.493 & 57.26 \\
3 & $1.27\times10^{-5}$ & 0.431 & 54.98 \\
4 & $1.45\times10^{-5}$ & 0.369 & 54.64 \\
\hline
\end{tabular}

\end{table}

\subsection{Acoustic Parameter Estimation}
\label{subsec:Acoustic Parameter Estimation Accuracy}
\vspace{-0.2cm}
Additional analyses such as key room acoustic parameters (T20, EDT and C50) were derived from both the target and predicted EDCs. The predictive performance was quantified using MAE, RMSE, and $R^2$ (coefficient of determination) for each parameter across all room configurations in the test dataset. Table \ref{tab:acoustic_param_accuracy} summarizes the error metrics. The model shows a good performance in estimating EDT with a MAE of \SI{0.033}{s} and an RMSE of \SI{0.042}{s} and T20 with MAE of \SI{0.023}{s} and an RMSE of \SI{0.029}{s} respectively. The high $R^2$ value of \SI{0.92}{} and \SI{0.97}{} for EDT and T20 respectively indicates that the predictions closely follow the trend of target values, confirming the model's ability to learn the early decay slope of the EDCs.

\begin{table}[t]
\centering
\caption{Performance of LSTM Model: MAE, RMSE, and coefficient of determination ($R^2$) computed between predicted and target EDCs}
\vspace{4pt} 
\begin{tabular}{@{}lccc@{}}
\toprule
\textbf{Metric} & \textbf{MAE}& \textbf{RMSE}& \textbf{R\textsuperscript{2}} \\
\midrule
EDT (s)& 0.033& 0.042& 0.92\\
T20 (s)& 0.023& 0.029& 0.97\\
C50 (dB)& 0.91& 2.53& 0.84\\
\end{tabular}
\label{tab:acoustic_param_accuracy}
\end{table}
\vspace{-0.25cm}
\begin{comment}
Figures \ref{fig:EDT} and \ref{fig:T20} compare the predicted vs. target values for each parameter. Across these graphs, a close alignment along the identity line ($y = x$) indicates strong agreement with the model. Particularly, the clustering around the ideal match line was the most prominent suggests that the model captured early reverberant characteristics effectively. The model shows a good performance in estimating EDT and  T20, with a low MAE of \SI{0.017}{s} and an RMSE of \SI{0.023}{s} and MAE of \SI{0.021}{s} and an RMSE of \SI{0.029}{s} respectively. The high $R^2$ value of \SI{0.987}{} and \SI{0.978}{} for EDT and T20 respectively indicates that the predictions closely follow the trend of target values, confirming the model's ability to learn the early decay slope of the EDCs.
    
\end{comment}

\begin{comment}

\begin{figure}
    \centering
    \vspace{-0.25cm}
    \includegraphics[width=1\linewidth]{EDT_comparison_6000_88200.png}
    \caption{Comparison of predicted and simulated $EDT$ values}
    \label{fig:EDT}
    %\vspace{-0.25cm}
\end{figure}

\begin{figure}
    %\vspace{-0.25cm}
    \centering
    \includegraphics[width=1\linewidth]{RT20_comparison_6000_88200.png}
    \caption{Comparison of predicted and simulated $T20$ values}
    \label{fig:T20}
    %\vspace{-0.25cm}
\end{figure}
\end{comment}

\subsection{Perceptual Validation of RIR Reconstruction}
\label{subsec:mushra}
We evaluated the perceptual plausibility of reconstructed RIRs generated from predicted EDC. From these predicted EDCs, RIRs were generated and subsequently compared against RIRs obtained from actual EDCs, which served as the perceptual reference. The reconstructed RIRs from both methods in section \ref{subsec:rirSynthesis}, were used to convolve four audios (\textit{Casta (music)}, \textit{Mix (Music+Speech)}, \textit{Male Speech}, and \textit{o444 (voice-based)}). Four different RIRs were selected with reverberation times ($T20 = 0.5 s, 1.34 s, 2.0 s$, and $3.0 s$). For each audio-reverberation combination, four stimulus conditions were generated: the \textit{reference} (RIR synthesized from the actual EDC), one anchors (\textit{Anchor35} \cite{schoeffler2018webmushra}, \textit{Random-Sign} (RIR synthesized from the target EDC using RS approach) and \textit{Random-Sign-Sticky} (RIR synthesized from the target EDC using RSS approach). This resulted in sixteen trials per participant, each trial containing four stimuli.

\begin{figure}
    \centering
    \vspace{-0.25cm}
    \includegraphics[width=1\linewidth]{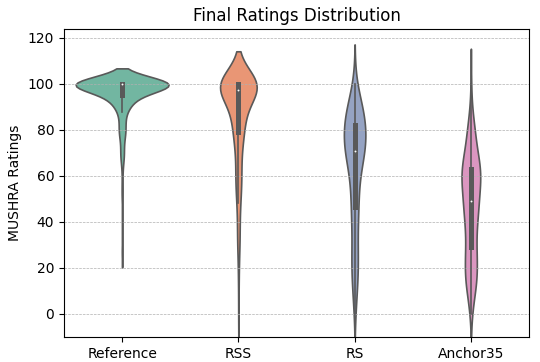}
    \caption{MUSHRA rating score: Refence, RSS (random sign sticky), RS (random sign) and Anchor35}
    \label{fig:mushra}
    \vspace{-0.25cm}
\end{figure}

A MUSHRA listening experiment was conducted following ITU-R BS.1534 recommendations. We used webmushra software \cite{schoeffler2018webmushra}. Twelve normal-hearing participants, all with prior experience in audio research or perceptual testing, rated the \textit{basic audio similarity} of each stimulus on a $0-100$ scale, where $100$ corresponded to the hidden reference and $0$ to the worst quality. Trials were presented in randomized order, and listeners underwent a training session to familiarize themselves with the interface and the range of anchor degradations. The results are show in the Fig \ref{fig:mushra} and Table \ref{tab:mushratable}.

\begin{table}[ht]
\centering
\caption{Summary statistics for different MUSHRA rating stimuli}
\vspace{4pt}
\label{tab:rating_stimulus}
\begin{tabular}{lrrrrr}
\toprule
Rating Stimulus & Mean & Std & Median & SEM & CI95 \\
\midrule
Reference            & 95.32 & 9.82  & 100.0 & 0.68 & 1.34 \\
RSS                  & 85.81 & 20.61 & 97.5  & 1.43 & 2.82 \\
RS                   & 63.17 & 24.93 & 71.0  & 1.73 & 3.41 \\
Anchor35             & 46.67 & 22.24 & 49.0  & 1.54 & 3.04 \\
\bottomrule
\end{tabular}
\label{tab:mushratable}
\end{table}

\vspace{-0.5cm}

\section{Summary and outlook}
\label{sec:Summary and outlook}
The proposed LSTM model predicts EDCs from room geometrical and material features, achieving low mean absolute error and root mean square error, and and high $R^2$ for EDT, T20 and C50 across unseen room configurations. RIRs reconstructed via the random sign–sticky method preserve temporal and spectral structure, including low-frequency coherence. These reconstructions were further validated through a MUSHRA listening experiment, showing mean ratings of $85.8$ for random sign-sticky compared to $95.3$ for the reference. The approach provides fast, scalable room acoustics prediction and auralization, suitable for interactive applications in VR and acoustic design. The source code, preprocessed dataset, and trained model used in this study are available at \cite{AMSgithub}.

\vfill
\pagebreak

\bibliographystyle{IEEEbib}
\bibliography{refs}

\end{document}